**Exploring Core-to-edge Transport in Aditya Tokamak by Oscillations Observed in the edge Radiation**

M. B. Chowdhuri, D. Raju, R. Manchanda, Vinay Kumar, Shankar Joisa, P. K. Atrey, C. V. S. Rao, R. Jha, R. Singh, P. Vasu and Aditya Team

*Institute for Plasma Research*
*Bhat, Gandhinagar – 382 428, India*

## Abstract:

Understanding of the transport in a Tokamak plasma is an important issue. Various mechanisms have been reported in the literature to relate the core phenomenon to edge phenomenon. Sawtooth and Mirnov oscillations caused by MHD instabilities are generally observed in Tokamak discharges. Observation of these effects in the visible radiation from outer edge may offer a possible means to understand the transport.

Oscillations in the visible radiation from outer region of the plasma have been observed during recent Aditya discharges. Percentage modulation of these oscillations vary with the Lines of Sight (LOS) of the chords and surfaces on which they terminate. This has been found in both the low frequency (~ 1 kHz) oscillations that seem to correlate with sawteething in SXR signals and the higher frequency (~ 10 kHz) oscillations that correlate well with Mirnov signals indicative of m/n = 2/1 mode rotation. This suggests that the extent to which the MHD instabilities in the central region of the plasma column are reflected in the edge radiation depends on the interaction of the plasma with the surface at the extremity of the LOS. The release of particle/ energy accompanying the MHD instabilities leads to a large influx of particles from such surfaces. Cross-bispectral analysis suggests that a mode (having frequency of ~ 20 kHz) is also generated due to the interaction of m/n =1/1 (~10 kHz, seen in SXR) and m/n = 2/1 (~10 kHz, seen in Mirnov, Visible & Microwave Interferometer signals). By possible selection rules, this mode seems to be a m/n = 3/2 mode. This mode is seen in Mirnov, Visible & Interferometer signals.

Behaviour of these oscillations on various LOS and their relation to SXR & Mirnov signals can lead to an understanding of the transport phenomenon. These observations and our interpretations will be presented.

e-mail: malay@ipr.res.in

## Introduction:

In this paper we report experimental observations on Aditya tokamak discharges, in which the sawteething activity in the core leads to periodic changes in the $H_\alpha$ signals with a surprisingly short delay of 200 microsec. These modulations are also observed in the ECE emissions from different radial locations between the core and the edge, with the delay periods progressively increasing outwards. These observations are strongly reminiscent of many observations in SXR and ECE emissions from other tokamak [1-4] and other magnetic confinement machines [5], which have lead to the postulation of a fast propagation of the energy released in a sawtooth [ST] crash, as a "heat pulse" (by which usually an electron temperature pulse is meant). In the light of available experimental evidence from Aditya, we have tried to infer the nature of the causal connection between the $H_\alpha$ modulation and sawteething in terms of heat pulse propagation.

In the first section of the following, the experimental details of the Aditya Tokamak and the parameters of discharges typical of the ones analyzed in this paper are given. In the next section all the observations that bear upon the possible link between ST oscillations in the core and modulations in emissivity of $H_\alpha$ in the edge, are described. This is followed by an analysis of the observations to evaluate their consistency with the fast propagation of energy.

## Aditya Tokamak:

The ADITYA [6] tokamak ($R = 75$ cm, $B_T = 0.75$ T, $a = 25$ cm with SS wall and a poloidal graphite limiter) has been producing ohmically heated circular plasma discharges. Typical parameter are: duration 70-80 ms, $I_P \sim 65\text{-}75$ kA, $n_e(0) \sim 1.5\text{x}10^{13}$ $cm^{-3}$, and $T_e(0) \sim 300$ eV. The central chord averaged electron density is measured by a microwave interferometer. The estimation of electron temperature is by the ratio method using soft X-ray emission from core. Bolometer collecting VUV radiations along a chord passing through center of the plasma gives an estimation of radiated power. The impurities line emissions and $H_\alpha$ emissions are routinely monitored by PMT-Interference filter based detectors and normal incidence VUV monochromator (NIM). A multichannel heterodyne radiometer detects the ECE emissions from different radial locations. A set of 15 Mirnov coils monitors poloidal magnetic field fluctuations. Fig. 1 shows the location of different diagnostics on Aditya tokamak.

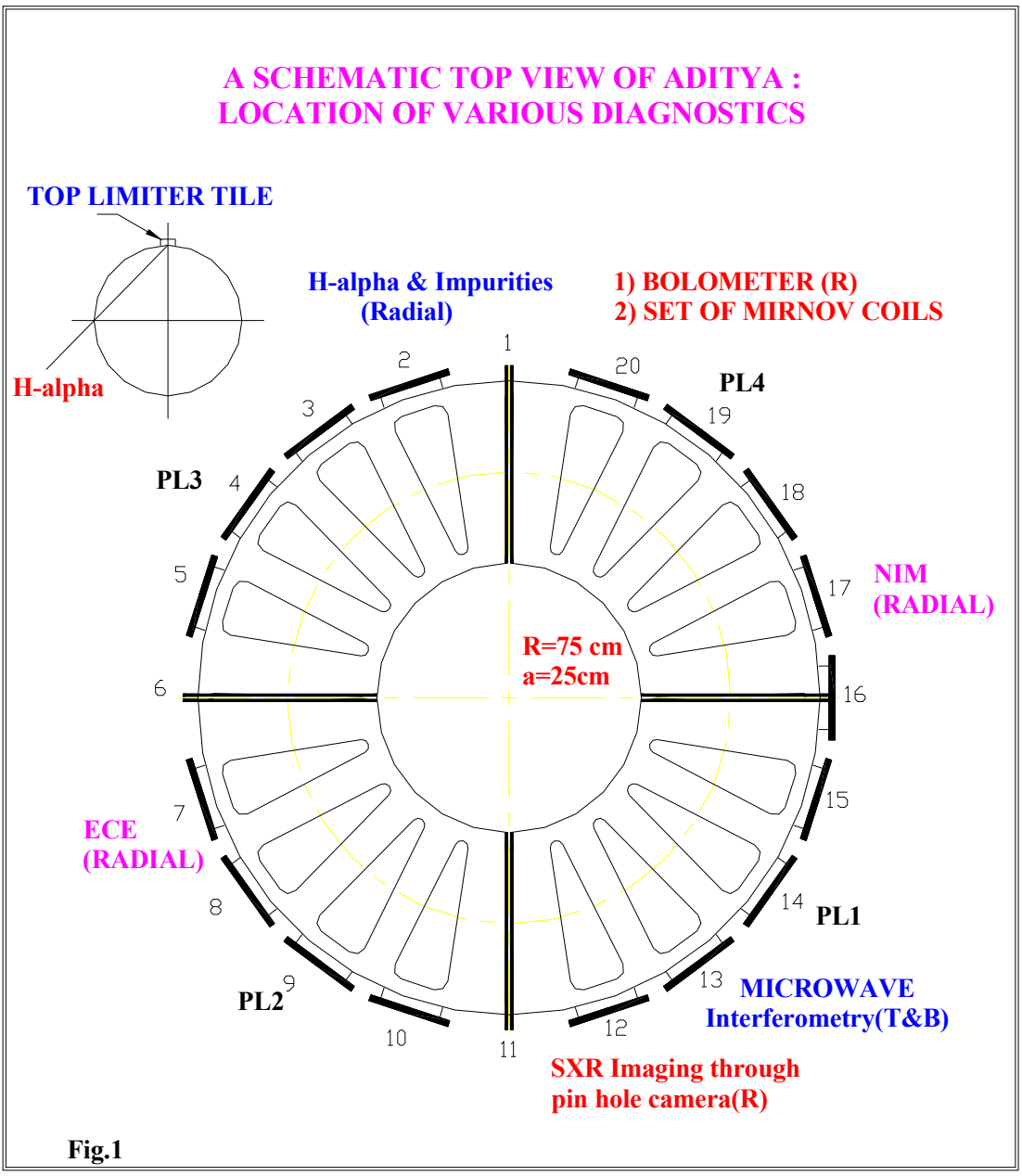

Fig.1

## Results:

Figure 2 shows Soft X-Ray (SXR) signals along a chord passing through plasma center and another along an off center chord. The sawtooth period is about 1 msec and the modulation is 30%. The core temperature (derived from the ratio of this signal to another filtered with Be-foil) is seen to change between 300 eV (pre crash) and 250 eV (post crash). In SXR signals the inverse ST pattern is seen usually along chords off centered by 6 cm and often in the 3 cm chord. From this we infer that the q = 1 radius is typically about 5 cm or smaller. The spatial resolutions of the SXR detectors and the small modulation in the weaker signals from the outer chords do not permit us to ascertain the inversion radius with higher precision for any particular shot. The above data leads to an estimate of ~ $2\times10^{19}$ eV (which of course critically depends on $r_{q=1}$) for the energy released from the core (q<1 region) during a ST crash.

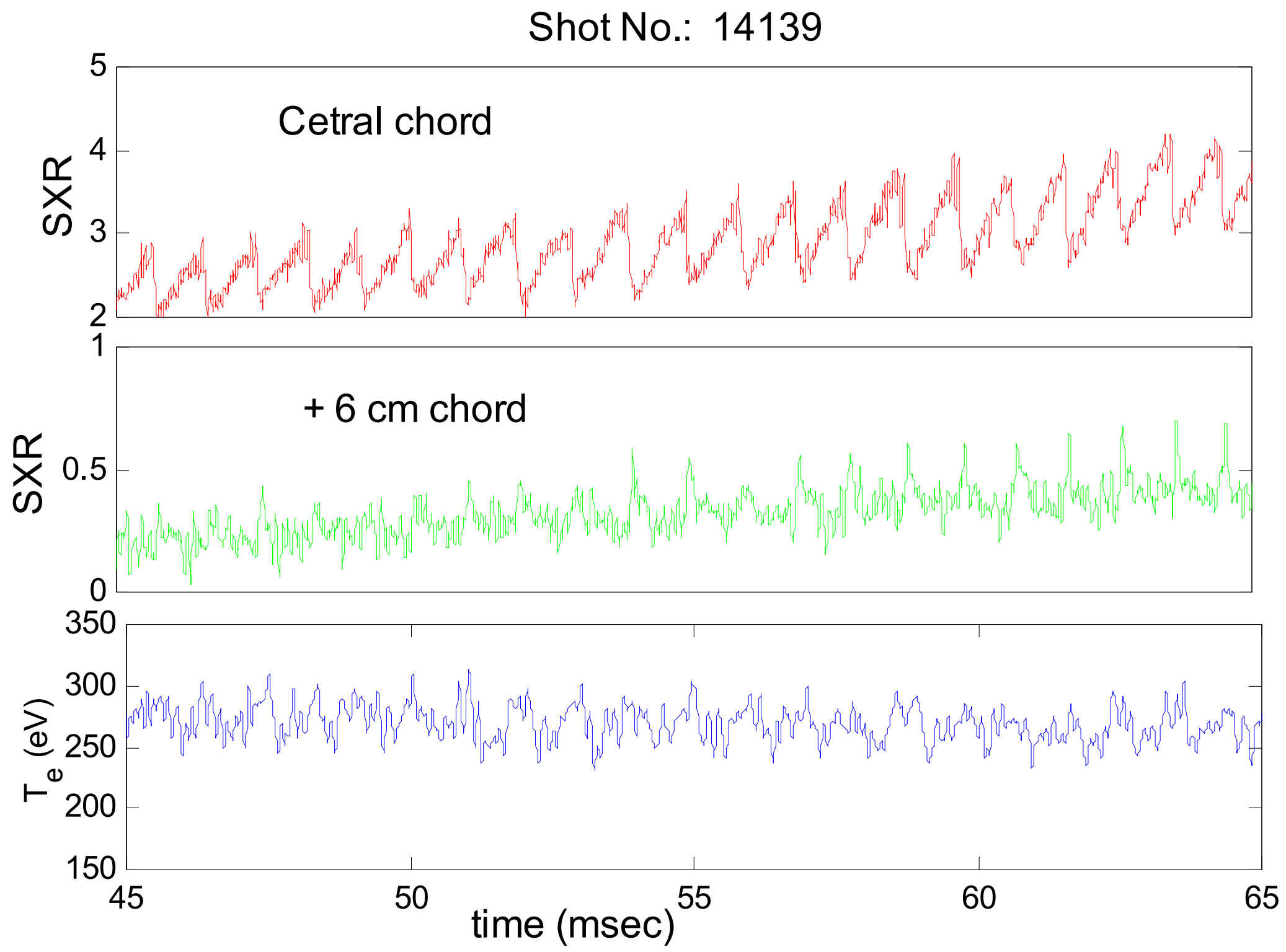

Fig. 2: SXR signal from different chords showing sawtooth and inverted sawtooth and derived electron temperature

Figure 3 shows the $H_\alpha$ signal monitored along an edge chord terminating on Aditya's top limiter tile (see inset of figure 1 for the line of sight of the $H_\alpha$ detector). Similar oscillations are also observed on the CIII signal along same chord. The oscillation corresponding to the ST (resembling more the inverse ST pattern) is evident in the signal itself. The cross correlations of the two signal yields an average value of ~ 210 microsec for the delay between ST crash and the $H_\alpha$ peak. Similar observations of ST oscillations modulating the $H_\alpha$ emissivity have been reported in many machines [3-5, 7].

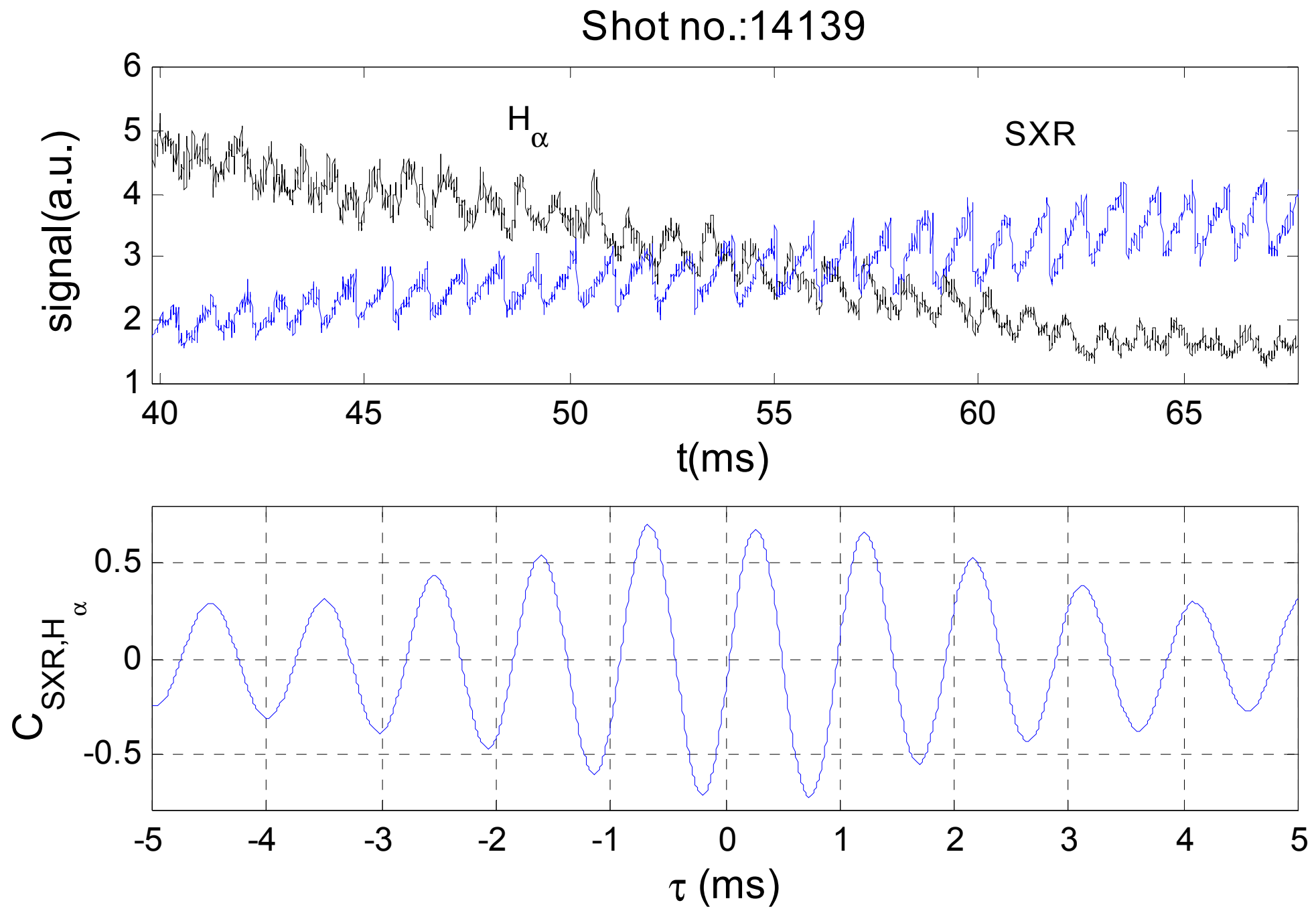

Fig. 3: Cross correlation between SXR and $H_\alpha$ shows 0.2 msec delay

Observation of similar delay has been reported for modulations in HXR detectors correlating to SXR sawteeth in PLT [8] and Pulsator [9]. In PLT the delay is ~ 2 msec, while in Pulsator it is 0.2 msec, close to what we observe as the delay between sawteeth crashes and $H_\alpha$ peak. The authors have investigated the possibility of ST induced de-confinement of runaway electrons impinging on the limiter and thus periodically altering the HXR emission from the target. It is conceivable that this would also result in bursts of hydrogen and impurity release and that this accounts for the modulations seen in $H_\alpha$ and CIII in our tokamak. But the HXR signal from the shots analyzed here does not show any ST modulations. In Aditya we have been able to observe such modulations in HXR signal (see fig 4) but only in shots, which are runaway dominated as inferred by the high level of HXR. The shots in the present campaign cannot be considered as runway discharges. So we shall not consider the possibility further, though ST instability itself generating runaways, which impinge on limiter (but any modulation in HXR being lower than our detection limit) cannot at present be ruled out. In the following we consider our observations in light of sawteeth induced energy release affecting the thermal electrons only.

In the JET tokamak [4] the ST crashes, during which $T_e$ (0) drops from 3.4 to 3.0 keV, ($\Delta T_e/T_e$ ~ 15%) with a periodicity of 100 msec, correlate to $H_\alpha$ peaks observed a few tens of millisecond later. The authors take this to be an indication of a heat pulse reaching the edge after a delay. In HT-7 tokamak [3] also ST (period ~ 5 msec, $\Delta T_e/T_e$ ~ 10%) leave their imprint on the $H_\alpha$ signals. In this paper also authors explain it as the effect of heat pulse propagating to edge. They take it to mean that the $H_\alpha$ signal is altered due to the heat pulse impinging on the wall enhancing the H atoms released from the surfaces. In other words, the enhancement in edge $H_\alpha$ emissivity is attributed to increased H atom flux. In our tokamak, the temperature variation $\Delta T_e$ in ST is nearly 50 eV and $\Delta T_e/T_e$ is quite comparable to that in ohmically heated plasma in the above machines. But the periodicity of ST (~ 1 msec) is smaller. We believe that it is unlikely that the energy released and then spread over the vessel surface area would significantly affect the H atom release. Also the surface thermal time constant involved would tend to smear out the release and destroy all coherence of $H_\alpha$ with the ST bursts.

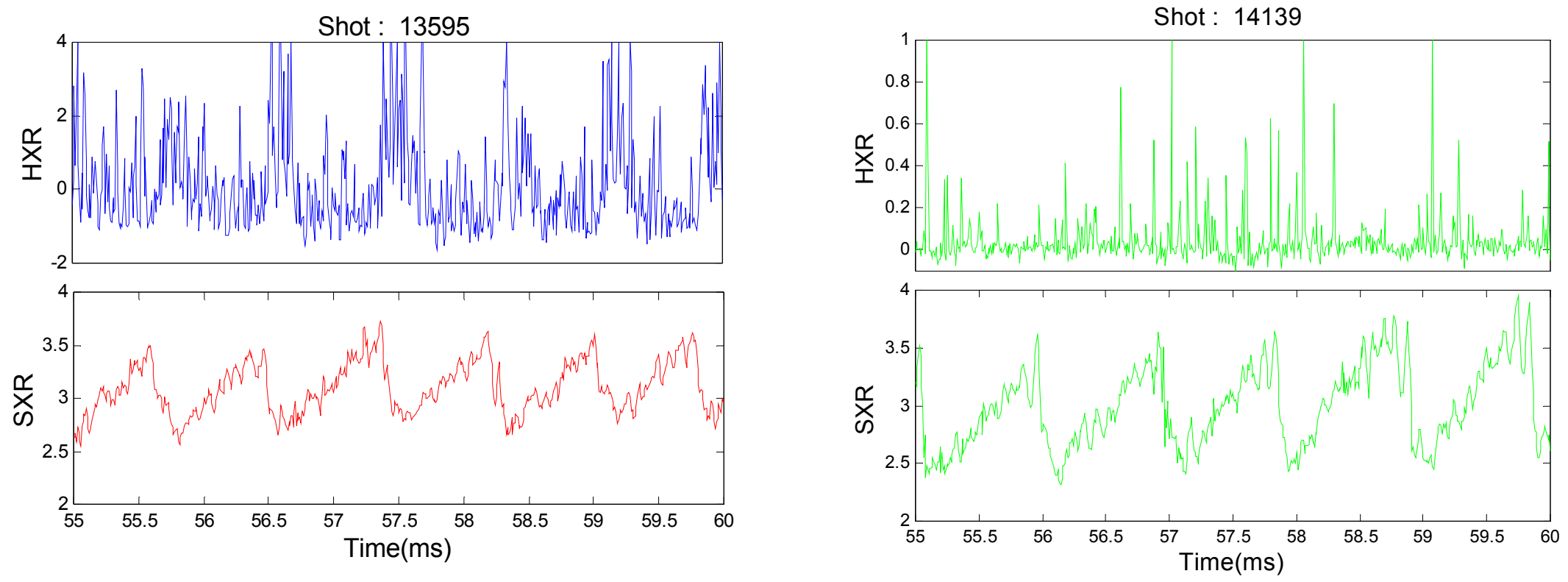

Fig.4: Modulation in HXR signal in different shots. Note the difference in HXR axes and absence of modulation in shot 14139.

We propose the following alternative mechanism to explain our observations. The arrival of the heat pulse (i.e. transient increase of electron temperature) at the edge momentarily alters the excitation rate. A temperature change $\Delta T_e$ in the local ambient temperature about 20 eV would result in an appreciable change in emissivity of $H_\alpha$, since rate of change of relative excitation coefficient $(1/X_e)*(dX_e/dT_e)$ ~ 0.03 $eV^{-1}$ at 20 eV [10]. In agreement with the above, ECE signal (presented later) from the edge, which is a measure of temperature, shows 33% - 40% modulations. The $\Delta T_e$ indicated by ECE measurement should result in relative intensity change in $H_\alpha$ of 20%, while the observed modulation is about 15%.

We are unable to observe ST related modulation in the line emissions of higher ionization species like OVIII, CV, NV from intermediate zones, at least not in the transitions that we use on Aditya to monitor these species. These are listed in Table 1. This is despite the fact that oscillation attributable to m = 2 mode rotations (and correlating well to Mirnov signals) are picked up routinely in these emissions (see fig. 5). This attests to the fact that the frequency response and S/N ratio of our detectors should have sufficed to record the sawteeth effects on the signals had they been present. It is estimated that for the weaker signal like OVII a modulation of <10%, and even smaller modulation for stronger signals like NV, would have been detected. HT-7 group [3] has also noted the absence of ST modulations in the impurities line emission. On the other hand, other groups [11] have reported such modulations, for example, in OVII at the core and CrI at the edge.

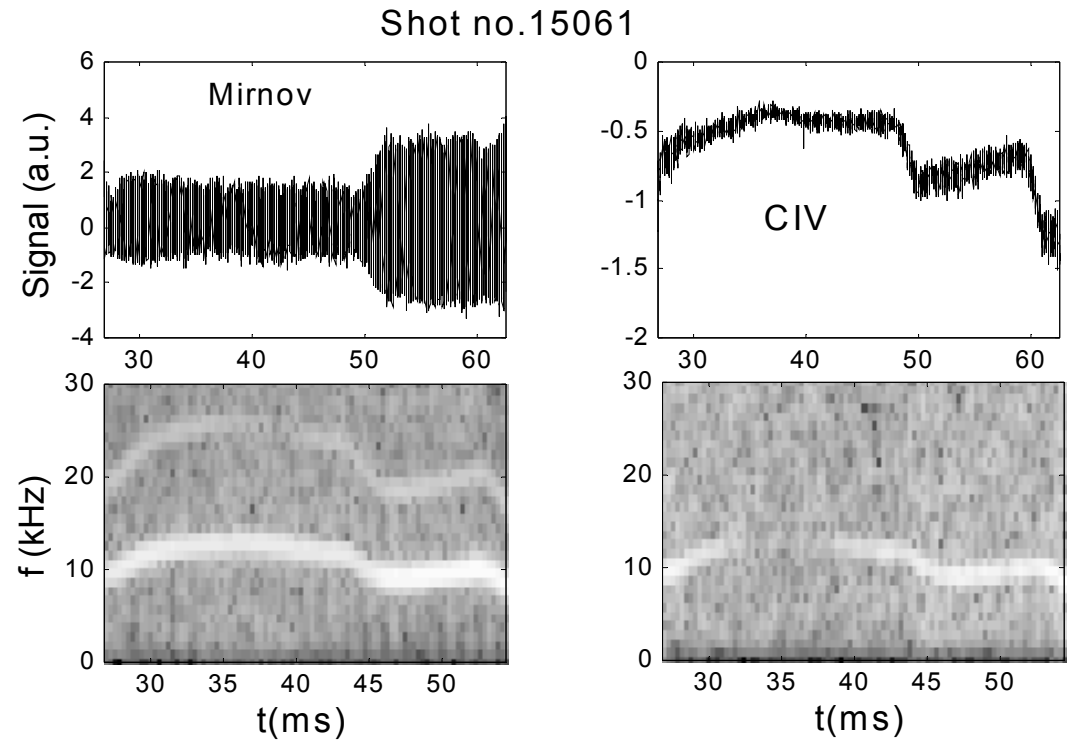

Fig. 5: CIV (1548 $A^0$) signal showing m=2 mhd oscillation

Table: 1

| Specie | I.P. (eV) | $\lambda$ (Å) | Transitions | Approx. upper & lower level energy (eV) | $1/X_e{*}(dX_e/dT_e)$ ($eV^{-1}$) [From ref. 10,12] |
|---|---|---|---|---|---|
| $H_\alpha$ | 13 | 6563 | $3p \rightarrow 2s$ | 12-10 | 0.05 @ 20 eV |
| CIV | 64 | 1548 | $1s^22p(^2P) \rightarrow 1s^22s(^2S)$ | 8-0 | 0.0021 @ 30 eV |
| NV | 97 | 1238 | $1s^2(^2P) \rightarrow 1s^2(^2S)$ | 10-0 | |
| CV | 393 | 2271 | $1s^22p(^3P) \rightarrow 1s^22s(^3S)$ | 304-300 | 0.0013 @ 200 eV |
| OVII | 871 | 1623 | $1s2p(^3P) \rightarrow 1s2s(^3S)$ | 568-561 | 0.0015 @ 400 eV |

This can be understood as follows. In the central regions of tokamak plasma, a particular ionization stage of specie is typically present in a shell where the temperature is about one half the ionization potential [13]. So we can expect that the emissivity for the particular transitions of the ion will be affected if the coefficient of excitation to the upper level of this transition is altered significantly by the change in temperature $\Delta T_e$ brought about by the heat pulse in the shell against the ambient temperature $T_e$. For the lines used, we note that the electron impact rate coefficients for excitation from ground level to the upper level of transitions are near their peak. The excitation coefficient data for CIV are shown in figure 6 for illustration. The relative variation in $X_e$ i.e. $1/X_e{*}(dX_e/dT_e){*}\Delta T_e$ (see table 1) is negligible as compared to that for $H_\alpha$. Even for a change of few tens of eV, the expected modulation would be hard to detect.

We conclude that the electron temperature change produced by the propagating heat pulse generated by ST crash in our discharges cannot result in appreciable change the emissivities of the lines presently monitored. In figure 6 we have also shown the excitation data for $1s^23d(^2D)$ level of CIV. Here, at the same shell temperatures of ~ 30 eV, the excitation rate coefficient is still quite sensitive to temperature. Lines originating from this level, for example 384 Å of CIV, would probably have shown ST related modulations. Similar transitions exist for other species as well at shorter wavelengths. In Aditya, a bolometer sensitive to all VUV wavelengths 10 Å - 2000 Å and viewing a central chord of the tokamak is available. The signal from the bolometer (see fig 7) does show modulations corresponding to the ST. However, the signature of individual rise and fall of sawteeth is not as well pronounced as in the case of individually resolved $H_\alpha$ and CIII signals though the presence of the ST frequency component is apparent. This is not surprising considering that the bolometer signal is not only a composite of all VUV lines but also of such radiations emanating from different shells.

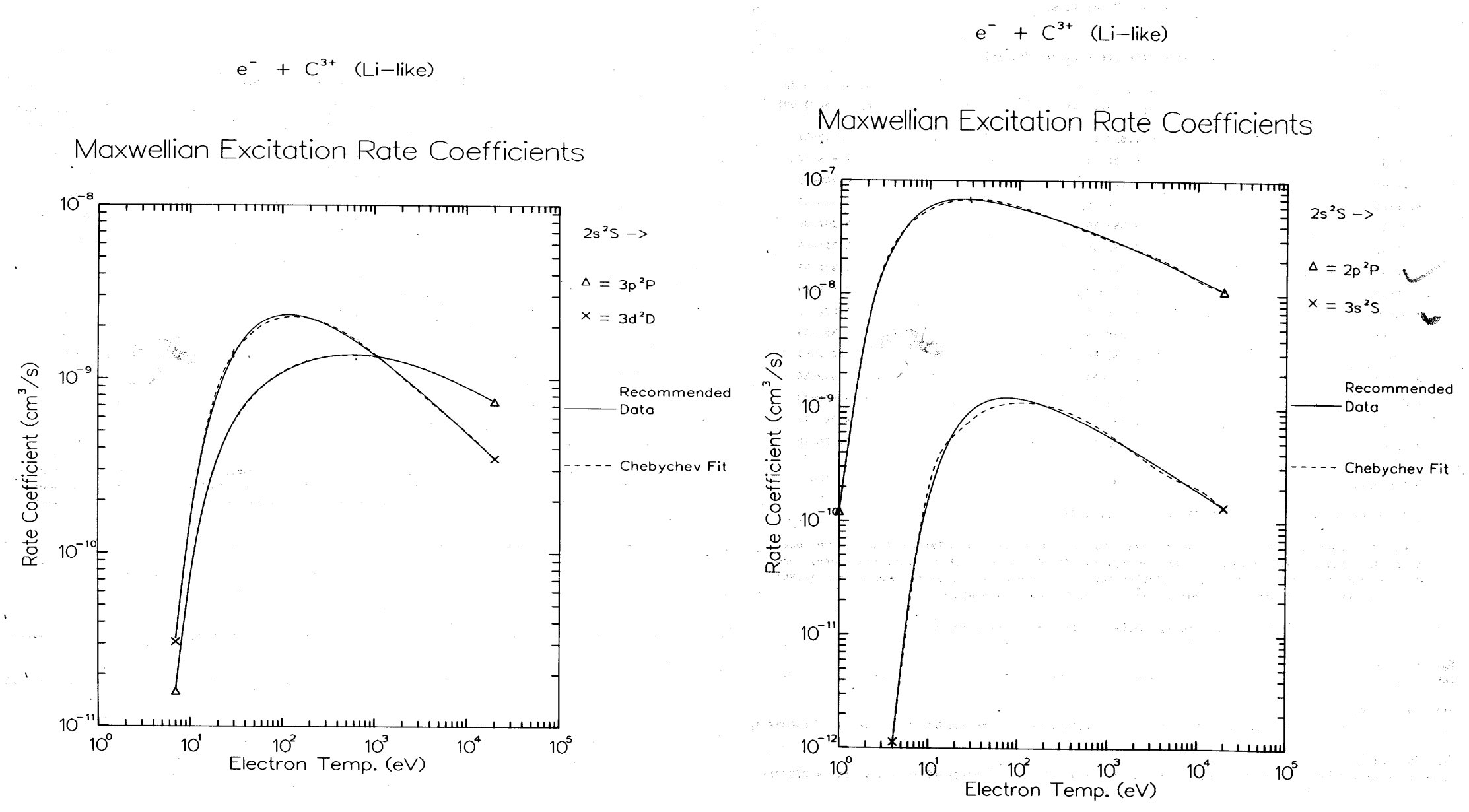

Fig: 6: Comparison of excitation rate coefficient populating the 2p($^2$P) and 3d($^2$D) of levels of CIV (reproduced from reference 12).

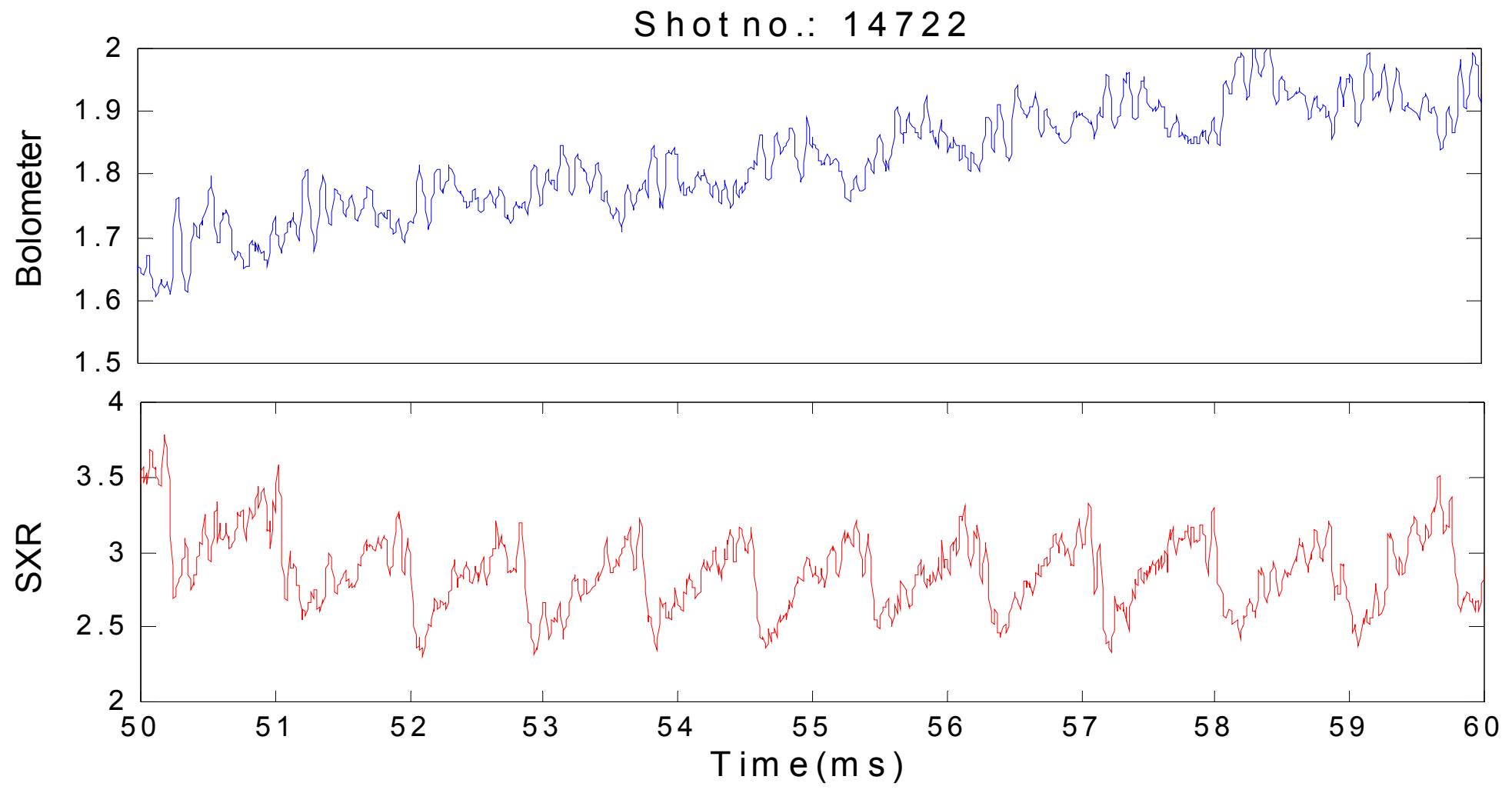

Fig. 7: Signature of sawtooth modulation on bolometer signal

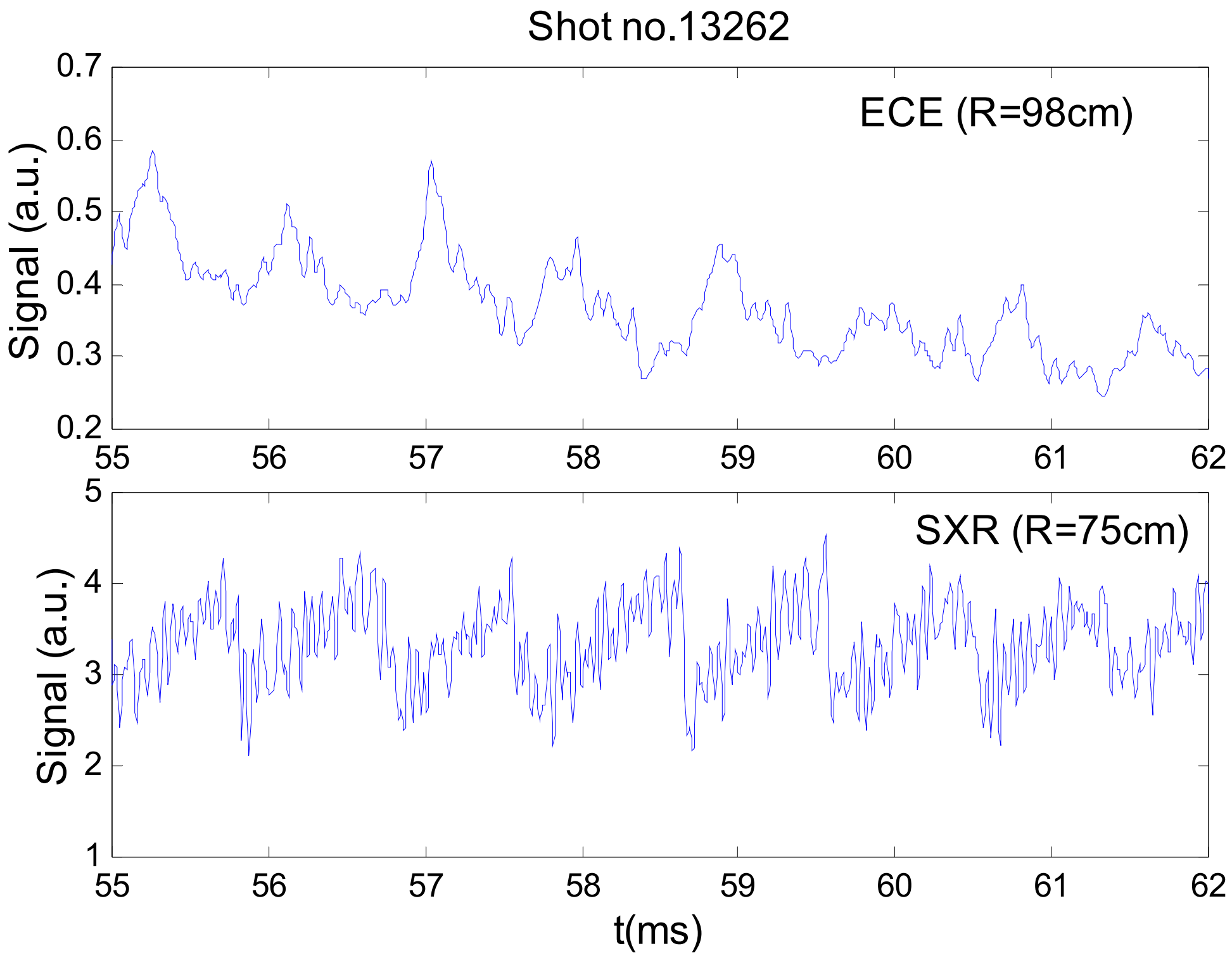

Fig.8a: Sawtooth induced modulation in edge ECE signal

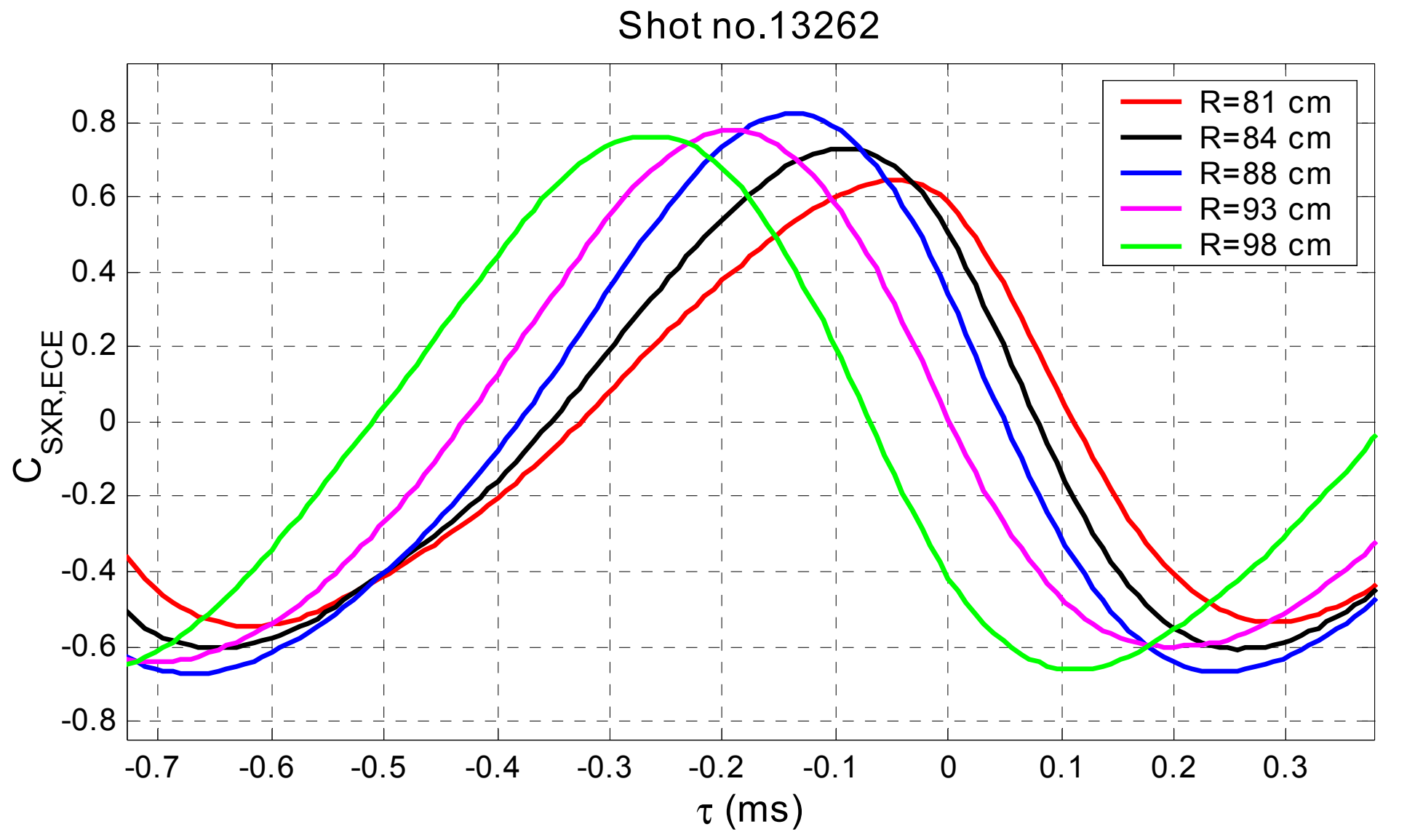

Fig. 8b: Cross correlation between SXR(R=75cm) and various ECE channels

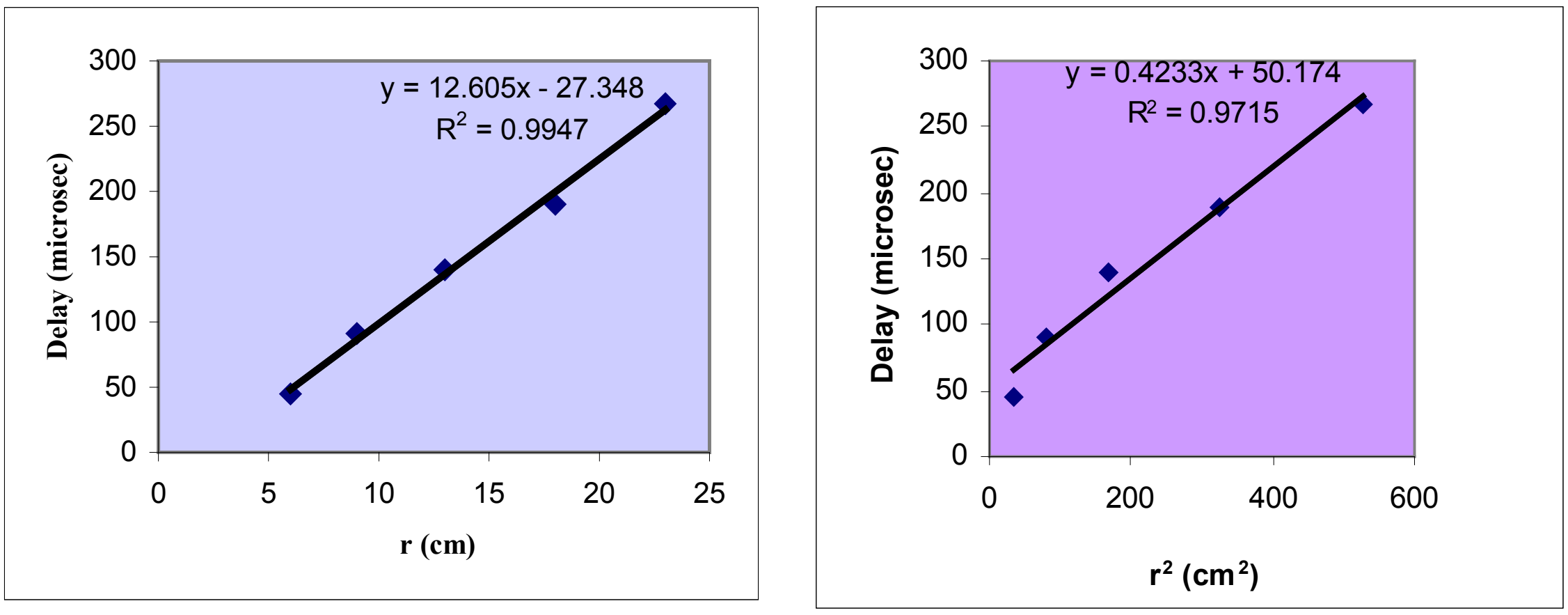

Fig. 9: Delay has been plotted against r and $r^2$

The available ECE emission data from the edge (r = 23 cm) as shown in figure 8a also clearly show the ST modulations. The delay between ST crash and the peak of ECE signal from different radial locations have been inferred by cross correlation analysis (fig 8b). The fact that ECE modulations at different radii show different delays is of course indicative of the transport of a ST induced energy pulse. In figure 9, we show the delay $\tau_d$ as a function of r and $r^2$, to seek whether the propagation is a ballistic (with a constant velocity) or a diffusive process. While the quality of linear fit to both is comparable, our data indicates a very rapid transport of energy from the core to edge with a ballistic velocity of 783 m/sec. Even if the propagation is viewed as purely diffusive, the inverse of the slope, which is the measure of the diffusivity, has a large value of ~233 $m^2$/sec.

## Conclusion:

Sawtooth related modulation has been observed in the edge radiation of Aditya plasma. We have argued that the modulation in $H_\alpha$ is due to change of excitation rate coefficient of transition upon the arrival of a heat pulse from the core. The specific lines emissions used to monitor the intermediate ionizations stages of impurities do not exhibit any modulation. This is also explained in terms of the relative insensitivity of their emission rate coefficients to small temperature changes during crash.
The delay between a ST crash and an $H_\alpha$ peak is very short and indicate an extremely fast propagation of the heat pulse generated during a sawtooth crash. The linear dependence of propagation delay time or distance suggests a ballistic component in the propagation.

## Acknowledgement:

Authors are grateful to Prof. P. K. Kaw for valuable discussions. We especially acknowledge Mr. H. B. Pandya for contributing to the paper with his analysis of the ECE results.